\documentclass[twocolumn,pra,showpacs,preprintnumbers,superscriptaddress,amsmath,amssymb,10pt,aps,floatfix]{revtex4}

\usepackage[utf8]{inputenc}
\usepackage[T1]{fontenc}

\usepackage{amsmath}
\usepackage{appendix}
\usepackage{graphicx}
\usepackage{mathrsfs}

\usepackage[caption=false]{subfig}

\begin{document}

\title{Out-of-time-order correlator in the quantum Rabi model}

\author{Aleksandrina V. Kirkova}
\affiliation{Department of Physics, St. Kliment Ohridski University of Sofia, James Bourchier 5 blvd, 1164 Sofia, Bulgaria}
\author{Diego Porras}
\affiliation{Institute of Fundamental Physics IFF-CSIC, Calle Serrano 113b, 28006 Madrid, Spain}
\author{Peter A. Ivanov}
\affiliation{Department of Physics, St. Kliment Ohridski University of Sofia, James Bourchier 5 blvd, 1164 Sofia, Bulgaria}

\begin{abstract}
We investigate signatures of chaos and equilibration in the quantum Rabi model, which exhibits a quantum phase transition when the ratio of the atomic level-splitting to bosonic frequency grows to infinity. We show that out-of-time-order correlator derived from the Loschmidt echo signal quickly saturates in the normal phase and reveals exponential growth in the superradiant phase which is associated with the onset of quantum chaos. Furthermore, we show that the effective time-averaged dimension of the quantum Rabi system can be large compared to the spin system size which leads to suppression of the temporal fluctuations and equilibration of the spin system.
\end{abstract}

\maketitle
\section{Introduction}

The quantum Rabi (QR) model is one of the simplest and most fundamental models describing quantum light-matter interaction. It consists of a single bosonic field mode and an effective spin system which interact via dipolar coupling \cite{Xie2017}. Various quantum-optical regimes of the QR model have been studied, including the ultra-strong coupling and deep strong coupling regimes, where the coupling strength is comparable to or larger than the bosonic mode frequency \cite{Pedernales2015,Lv2018}. Recently, it was shown that the QR model exhibits a finite-size quantum phase transition when the ratio of level-splitting $\Delta$ to bosonic frequency $\omega$ grows to infinity $\eta = \Delta / \omega \rightarrow \infty$ \cite{Hwang2015}. The latter corresponds to the classical oscillator limit $\omega \rightarrow 0 $ that also unveils a finite-size criticality in its generalized counterpart of $N$ two-level systems, the Dicke model \cite{Bakemeier2012}. The second-order quantum phase transition in the QR model occurs at a critical spin-boson interaction strength $g = g_{\rm c}$ between a normal $g < g_{\rm c}$ and a superradiant phase $g > g_{\rm c}$. 
The recent experimental realization of such a quantum phase transition in a trapped-ion system opened fascinating prospects for exploring critical behaviour in finite-size quantum optical systems \cite{Cai2021}.

Critical behaviour in quantum many-body systems has been associated with the onset of chaos \cite{Emary2003a,Emary2003,Rey2019,Perez2011,Georgeot1998}. 
In light of this we investigate signatures of chaos in the QR model as we approach the effective thermodynamic limit $\eta \rightarrow \infty$. 
One such measure is the nearest-neighbour level-spacing distribution of the Hamiltonian eigenenergies. While for non-chaotic systems we expect a Poissonian distribution \cite{BerryTabor1977}, the onset of chaos is associated with a crossover to Wigner-Dyson statistics, as described by Random Matrix Theory \cite{DAlessio2016}. We show that neither of these distributions is observed in the QR model, due to it being finite-size \cite{Kus1984}, however, the spectrum exhibits level-crossings in the normal phase and level-repulsions in the superradiant phase, the latter being associated with chaotic behaviour of non-integrable systems. 

Furthermore, we use a double commutator out-of-time-order correlation function (OTOC) which measures the scrambling of quantum information across the system's degrees of freedom \cite{Swingle2018}. The OTOC is presented as an indicator of quantum chaos, with its growth rate being associated with the classical Lyapunov exponent \cite{Shenker2014,Bohrdt2017,Shen2017}. Moreover, the OTOC has been measured in a system of trapped ions \cite{Garttner2017,Landsman2019,Joshi2020,Green2021} and in a nuclear magnetic resonance quantum simulator \cite{Li2017}. Recently, the thermally averaged OTOC with infinite temperature has been studied in quantum Rabi and Dicke models \cite{Sun2019}. Here we explore the OTOC derived from the Loschmidt echo signal \cite{Schmitt2019} to study the variance of an observable under imperfect time reversal. We show that in the normal phase the OTOC quickly saturates to a value independent of $\eta$. In the superradiant phase, however, it displays \emph{exponential growth} which becomes larger as $\eta$ is increased and allows for numerical extraction of the Lyuapunov exponent $\lambda_Q (g,\omega,\Delta)$ \cite{Maldacena2016, Knap2017,Carlos2019, Keselman2021}. We find that the relation $\lambda_Q t^* \sim \log\eta$, that is characteristic of chaotic systems with a classical limit of $1 / \eta \sim \omega \rightarrow 0$ \cite{Rammensee2018}, holds for the QR model, with $t^*  (g,\omega,\Delta)$ being the saturation time of the OTOC. Similarly, $\lambda_Q t^* \sim \log N $ has been proven to hold for a variety of quantum many-body systems \cite{Susskind2008,Rey2019,Chen2018,Gharibyan2019}. Moreover, we find similar exponential growth of the OTOC in other non-integrable critical quantum systems such as perturbed QR model and quantum Jahn Teller (QJT) model which indicates that the onset of chaos is closely related to the existence of a finite-size quantum phase transition.

Finally, we investigate the connection to equilibration and thermalization. We show that the long-time average of observables in the QR model relaxes to a value solely determined by the initial energy. We show that the observables of the QR model don't thermalize in general. However, we find a regime in which the effective dimension of the time-averaged density operator is larger than the spin system dimension which drives the spin system towards equilibrium \cite{Linden2009,Gogolin2011}.  

The paper is organized as follows: In Sec. \ref{QR} we introduce the QR model which exhibits a finite-size quantum phase transition. We investigate the nearest-neighbour level-spacing distribution and show that neither Poissonian nor Wigner-Dyson statistics is observed. In Sec. \ref{chaos} we discuss the fidelity out-of-time correlator as a measure of chaos in our model. We observe an exponential growth of the fidelity out-of-time correlator in the superradiant phase which is characterized by a quantum Lyapunov exponent and saturation time. In Sec. \ref{Equilibration} we show that effective dimension of the time average density operator can be sufficiently large such that the single bosonic degree-of-freedom acts as a bath coupled to the spin system. This leads to suppression of the temporal fluctuations and equilibration of the spin system. Finally, the conclusions are presented in Sec. \ref{SM}.
\begin{figure}[t]
\centering
\includegraphics[width=0.47\textwidth]{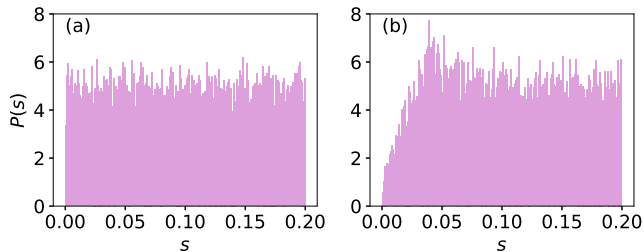} 
\caption{{Distribution $p(s)$ of the nearest-neighbor energy spacing $s_{n}=E_{n+1}-E_{n}$ for the QR model with Hamiltonian (\ref{RabiHamiltonian}). (a)  Normal phase with spin-boson copling $g=3$. (b) Superradiant phase with $g=7$. The other parameters are set to $\eta = 200$ with critical coupling $g_{\rm c}=5$. The bosonic Hilbert space is truncated at $n_{\rm max}=120000$.}}
\label{fig:rabi_levelspacing}
\end{figure}

\section{Quantum Rabi Model}\label{QR}
\subsection{Finite size quantum phase transition}
The QR Hamiltonian is given by
\begin{equation}
\hat{H}_{\rm QR}= \omega \hat{a}^\dag \hat{a} + \frac{\Delta}{2} \sigma_z + g \sigma_x (\hat{a}^\dag + \hat{a}), \label{RabiHamiltonian}
\end{equation}
where $\Delta$ is the level-splitting of the two-level system and $\hat{a}^\dag$, $\hat{a}$ are respectively the creation and annihilation operators of the bosonic mode, corresponding to an oscillator with frequency $\omega$. The coupling $g$ characterizes the strength of the dipolar spin-boson interaction. The QR model exhibits a finite-size quantum phase transition at the critical coupling $g_{\rm c} = \sqrt{\Delta\omega} / 2$ in the effective thermodynamic limit $\eta\rightarrow \infty$. The two phases of the system are a normal phase for $g<g_{\rm c}$ characterized by zero mean-field bosonic excitations and polarized spin along the $z$-axis, and a superradiant phase for $g > g_{\rm c}$ with non-zero magnetization along the $x$-axis and a macroscopically excited bosonic state \cite{Hwang2015,Cai2021}. 

\begin{figure}[tp]

    {\includegraphics[width=0.45\textwidth]{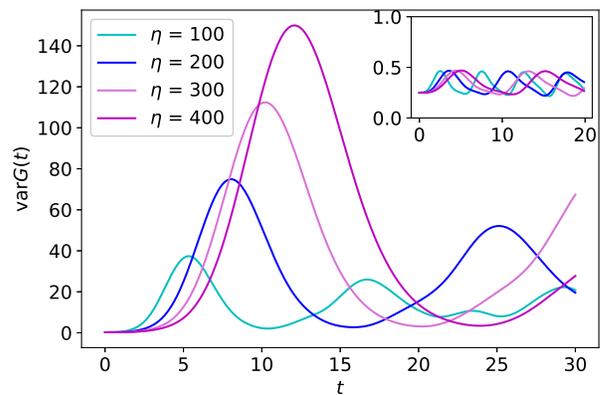}}
  \caption{Exact time-evolution of FOTOC for the QR model with Hamiltonian (\ref{RabiHamiltonian}). We set $g_{\rm c}=5$, and vary $\eta$. In normal phase with $g = 4$ the FOTOC oscillates with amplitude independent of $\eta$ (inset). In the superradiant phase with $g=7$ the FOTOC exponentially grows with quantum Lyapunov exponent $\lambda_{Q}(g,\omega,\Delta)$ and saturation time $t^{*}$. The initial state is $\left|\psi_{0}\right\rangle=\left|+,0\right\rangle$.}
  {\label{fig:fotoc}}
  \end{figure}
  
\subsection{Level-spacing distribution} 
Usually the cross-over between integrable and chaotic behaviour in quantum systems is related to the change of energy level statistics  from Poissonian $p_{\rm P}(s)=e^{-s}$ to the Wigner-Dyson distribution $p_{\rm WD}(s)=(\pi s/2)e^{-\pi s^{2}/2}$, which in random-matrix theory describes a chaotic system \cite{DAlessio2016,Gogolin2016}. In the core of this method lies the observation of level crossing for integrable systems and level repulsion for chaotic ones.

In order to consider the level-spacing statistics of the QR model one needs to first unfold the energy spectrum, so that the resulting distribution includes only transitions within a subspace of states that is invariant under the parity transformation, see Appendix \ref{QRLSD} for more details. In Fig. \ref{fig:rabi_levelspacing} we show the level statistics distribution for the QR Hamiltonian (\ref{RabiHamiltonian}). Although the level statistics distribution is neither Wigner-Dyson nor Poissonian, if we focus on small scales for the energy difference $s$, one can see that the QR model indeed exhibits level crossing in the normal phase ($g<g_{\rm c}$) and level-repulsions in the superradiant phase ($g>g_{\rm c}$). The QR Hamiltonian was shown to have a regular spectrum and was deemed integrable in \cite{Braak2011}, however, the spacing between adjacent eigenenergies is dependent on $\omega$, which in our effective thermodynamic limit tends to zero, thus indicating possible level-clustering as long as $\eta \rightarrow \infty$.


Furthermore, we investigate the level spacing distribution in the QJT model with Hamiltonian $\hat{H}_{\rm JT}=\hat{H}_{\rm QR}+\hat{H}_{b}$, where $\hat{H}_{b}=\omega \hat{b}^{\dag}\hat{b}+g\sigma_{y}(\hat{b}^{\dag}+\hat{b})$ which describes the U(1) symmetric interaction between a single spin and two bosonic modes \cite{Porras2012}. To the best of our knowledge the QJT model is not integrable. Similarly to the QR model, the QJT model exhibits a finite-size quantum phase transition in the limit $\eta\rightarrow\infty$ between a normal phase and a U(1) symmetry-broken supperadiant phase, see Appendix \ref{QJTM}. We observe neither Poissonian nor Wigner-Dyson nearest-neighbourgh distribution in both phases of the QJT system. However, focusing on a smaller scale for the energy difference, one can see that level-crossings are present in the normal phase, and level-repulsions in the supperradiant phase.

\section{Fidelity out-of-time-order correlators}\label{chaos}  
To further investigate signatures of chaos in QR model, we employ out-of-time-order correlation functions (OTOCs) 
\begin{equation}
F(t) = \langle \hat{W}^\dag (t) \hat{V}^\dag \hat{W}(t) \hat{V} \rangle,
\end{equation}
where the angular brackets denote averaging over the initial state $\left|\psi_{0}\right\rangle$. The OTOCs quantify the degree of non-commutativity in time between two initally ($t=0$) commuting operators $[\hat{W}, \hat{V}] = 0 $, whose time-evolution is governed by the system Hamiltonian as $\hat{W}(t) = e^{i\hat{H}t} \hat{W} e^{-i\hat{H}t}$. Moreover, it can be regarded as a natural extension of the idea of classical chaos via the correspondence between the phase space Poisson brackets and the commutator in quantum mechanics since $ 1 - \Re [F(t)] = \langle [\hat{V}^\dag, \hat{W}^\dag (t) ][\hat{W}(t),\hat{V}]\rangle/2 \sim e^{\lambda_Q t}$, where $\lambda_Q$ is a quantum Lyuapunov exponent, associated with the onset of chaos. In the following we choose $F(t)$ to be a fidelity OTOC (FOTOC) with the condition that the initial state $\left|\psi_{0}\right\rangle$ is an eigenstate of $\hat{V}$ and $\hat{W}_G = e^{i\delta \phi \hat{G}}$ for a Hermitian operator $\hat{G}$, where $\delta \phi$ is a small perturbation. Such a choice has been considered for studying the irreversibility of the dynamics in the Sachdev-Ye-Kitaev model due to imperfect time reversal \cite{Schmitt2019}, and for quantifying scrambling and quantum chaos in the Dicke model \cite{Rey2019}. We choose $\hat{V}$ to be a projector on the initial state $\hat{V} = \hat{\rho}(0)=|\psi_{0}\rangle\langle\psi_{0}|$ where $\left|\psi_{0}\right\rangle=\left|+,0\right\rangle$ ($\sigma_{x}\left|+,0\right\rangle=\left|+,0\right\rangle$). Note that alternatively one can set $\hat{V}=\sigma_{x}$, see Appendix \ref{echo}. Since $\delta \phi$ is a small perturbation one can expand the FOTOC $F_G (t) = \langle \hat{W}_G^\dag (t) \hat{\rho} (0) \hat{W}_G (t) \hat{\rho}(0) \rangle $ in power series of $\delta \phi$ which yields

\begin{equation}
1 - F_G(t) = \delta \phi^2 ( \langle \hat{G}^2 (t) \rangle - \langle \hat{G} (t) \rangle ^2 ) = \delta \phi^2 {\rm var} \hat{G}(t).\label{Gexpansion}
\end{equation}

\begin{figure}
\centering
\includegraphics[width=0.45\textwidth]{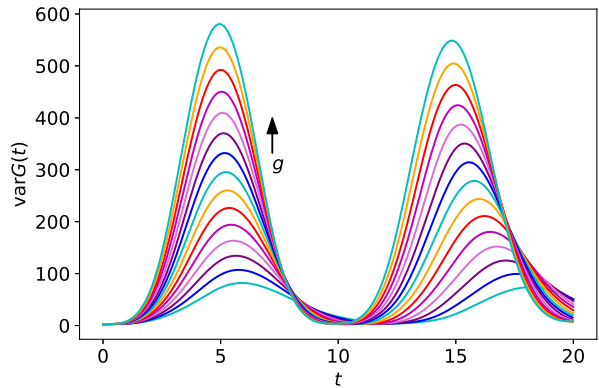}
\caption{Time evolution of the FOTOC for the QR model from the initial state $\left|\psi_{0}\right\rangle=\left|+,5 \right\rangle$. The coupling is varied from $g = 7$ to $g = 14$ in steps of $0.5$, $\eta = 200$.
\label{fig:4}}
\end{figure}

In Fig. \ref{fig:fotoc} we plot the the variance of $\hat{G} = (\hat{a}^\dag + \hat{a})/ 2$. We observe a clearly distinguishable difference in the behaviour of FOTOC in the two quantum phases. In the normal phase ($g<g_{\rm c}$) the FOTOC oscillates with an amplitude independent of $\eta$, see Fig. \ref{fig:fotoc} (inset). In the superradiant phase ($g>g_{\rm c}$) we observe exponential growth of the FOTOC in the beginning of the time evolution, which is associated with the onset of quantum chaos via the relation (\ref{Gexpansion}).
\begin{figure}[tp]
\centering
\includegraphics[width=0.45\textwidth]{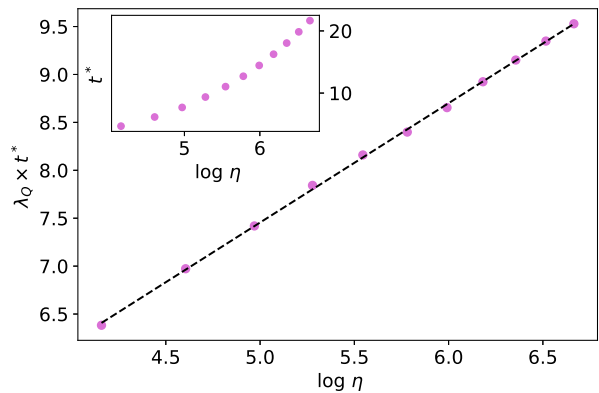}
\caption{Lyapunov exponent times the scrambling time $\lambda_Q(g,\omega,\Delta) t^*$ as a function of $\eta$. The parameters are set to $g = 6$, $g_{\rm c} = 5$. The relation is well approximated by a logarithmic function $\lambda_Q (g,\omega,\Delta) t^* \sim \log \eta$ (dashed line). (Inset) Scrambling time $t^*$ as a function of $\eta$ in the superradiant phase.
\label{fig:lambdaQ-fit}}
\end{figure}
The exponential growth is observed after a short time of slow dynamics with no perceivable growth of the FOTOC. From here we can extract the quantum Lyapunov exponent $(1 - F_G(t))/\delta\phi^2 \sim e^{\lambda_Q (g,\omega,\Delta) t}$. We observe that as $\eta$ increases the FOTOC grows larger and reaches its maximal value at the scrambling time $t^{*}$, beyond which any initial local information about the system is globally spread among its degrees of freedom. After the scrambling time $t^*$ the FOTOC displays oscillatory behaviour characterized by periodically occurring maximal saturation, see Appendix \ref{long}. In Fig. \ref{fig:lambdaQ-fit}(inset) we show the exact result for the scrambling time $t^{*}$ as a function of the parameter $\eta$. We find that $t^{*}$ behaves as $t^{*}\sim a\log(\eta)+b\log^{2}(\eta)$ with $a$ and $b$ being fit parameters. Moreover, we find the relation $\lambda_Q (g,\omega,\Delta) t^* \sim \log\eta$ as shown in Fig. \ref{fig:lambdaQ-fit}. Finally, we note that the FOTOC and the Lyapunov exponent increase with $g$, while the saturation time $t^{*}$ stays nearly constant which makes the QR system more chaotic for stronger spin-boson interaction, see Fig. \ref{fig:4}.

To expand on the connection between integrability and chaos, we turn to the QJT model and the perturbed QR Hamiltonian $\hat{H}_{\rm QR}^{(\lambda)}= \hat{H}_{\rm QR} + \lambda\sigma_x$, which is solvable, yet non-integrable due to the broken parity symmetry \cite{Braak2011}. We find that the FOTOC behaves very similarly in both of these models (albeit not for all parameter regimes of $\hat{H}_{\rm QR}^{(\lambda)}$), displaying initial exponential growth up to the saturation time, see Appendix \ref{QJTM}. Therefore, the signatures of chaos of these models are closely related to the existence of the finite-size quantum phase transition rather than their integrability.

\section{Equilibration}\label{Equilibration} 

\begin{figure}[tp]
\centering
\includegraphics[width=0.45\textwidth]{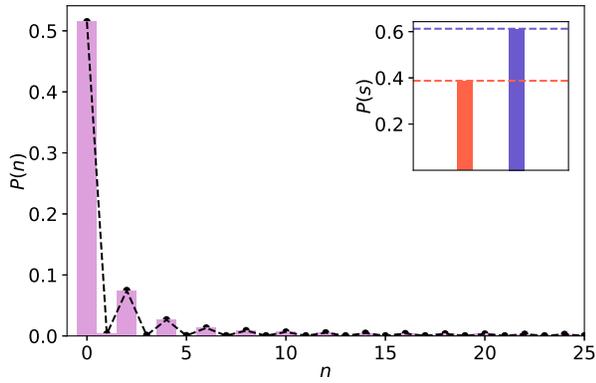}
\caption{Numerical result for the time-average mean boson occupation distribution $P(n)$ (pink bars) compared with its diagonal ensemble average with density operator $\hat{\rho}_{\rm DE}$ (dashed line). (Inset) The time-average spin population $P(\downarrow)$ (red bar) and $P(\uparrow)$ (blue bar) compared with its diagonal ensemble average (dashed lines). The parameters are set to $g = 10$, and $\eta = 200$.
\label{fig:fockstatesDE}}
\end{figure}

The emergence of quantum statistical mechanics in a closed system has been connected to chaotic behaviour, therefore investigation of equilibration and thermalization is a natural continuation of our discussion \cite{DAlessio2016}. The \emph{Eigenstate Thermalization Hypothesis} (ETH) \cite{Srednicki1994, Deutsch1991} states that the expectation value of a thermalizing observable in a Hamiltonian eigenstate is equal to the microcanonical prediction for that observable at the corresponding eigenenergy.
As our system exhibits temporal fluctuations, we focus on the long-time average of observables, rather that their true value. Given an initial state $|\psi_0 \rangle = \sum_{\alpha} c_{k} |E_{k}\rangle$, where $\hat{H}|E_{k}\rangle = E_k |E_k\rangle $ and $c_k = \langle E_k|\psi_0 \rangle$, evolving under the system Hamiltonian as  $|\psi (t) \rangle = e^{-i\hat{H}t} | \psi_0 \rangle = \sum_{k} c_k e^{-iE_k t} |E_k\rangle $, the long-time average of an observable $\hat{O}$ reads

\begin{equation}
    \bar{\langle {O} \rangle} = \sum_{k} |c_k|^2 O_{kk} = {\rm Tr}[\hat{\rho}_{\rm DE} \hat{O}], \label{longtimeavg}
\end{equation}
where $\hat{\rho}_{\rm DE}$ is the density matrix of the so-called diagonal ensemble (DE) $\hat{\rho}_{\rm DE} = \sum_{k} |c_k|^2 |E_k\rangle \langle E_k| $ and $O_{kk} = \langle E_k | \hat{O} | E_k \rangle$. The statement of the ETH translates to the fact that the predictions for $\bar{\langle {O} \rangle}$ given by the diagonal and microcanonical ensemble (ME) at energy $E_0 = \langle \psi_0 | \hat{H} | \psi_0 \rangle $ coincide, where the latter is given by
\begin{equation}
{\langle \hat{O} \rangle}_{\rm ME} (E_0) = {\rm Tr}[\hat{\rho}_{\rm ME} \hat{O}]= \frac{1}{\mathcal{N}} \sum_{k : |E_k - E_0| <\delta E} O_{kk}.
\end{equation}
The sum runs through the $\mathcal{N}$ eigenstates of $\hat{H}$ that are inside an energy shell of width $2\delta E$ around $E_0$.

\begin{figure}[tp]
 \centering
 \includegraphics[width=0.47\textwidth]{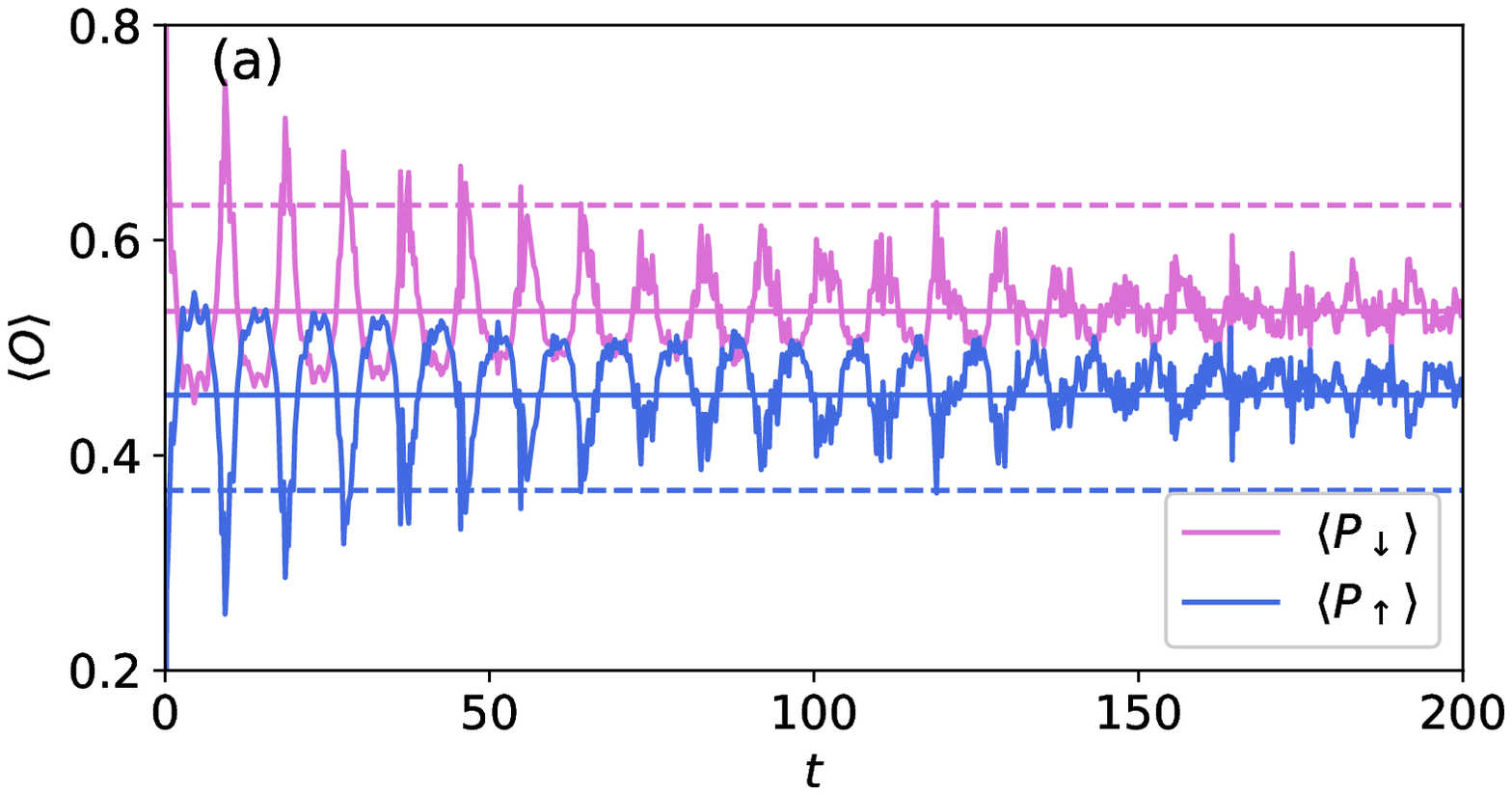}
  \includegraphics[width=0.22\textwidth]{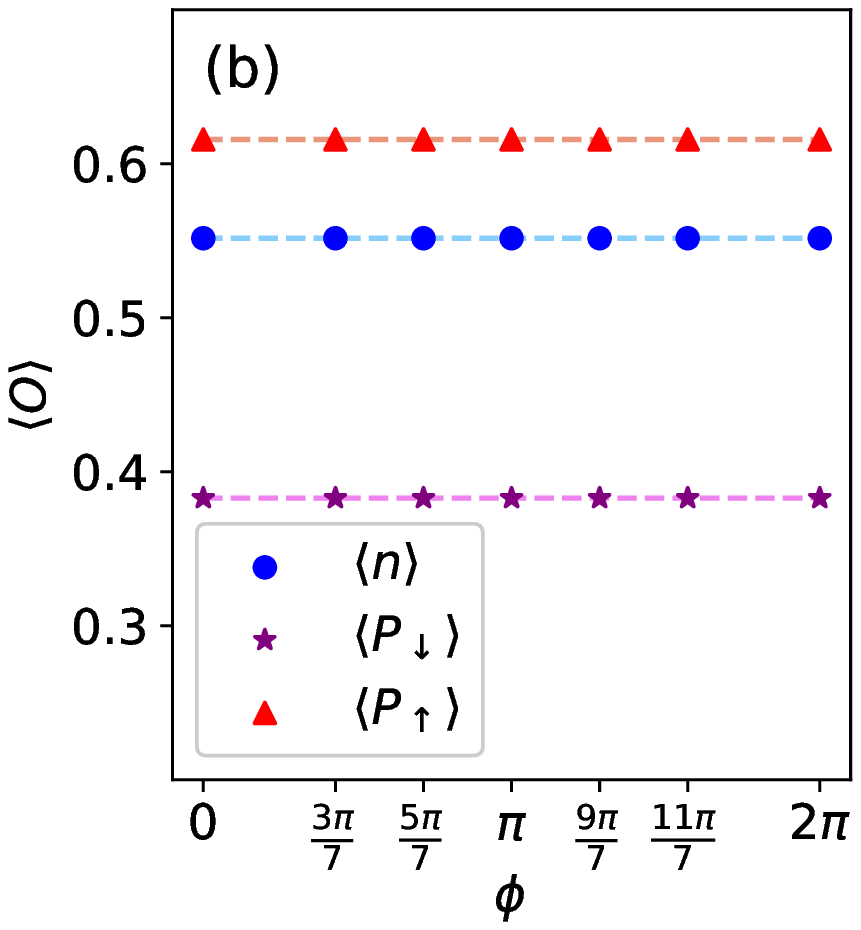}
  \includegraphics[width=0.225\textwidth]{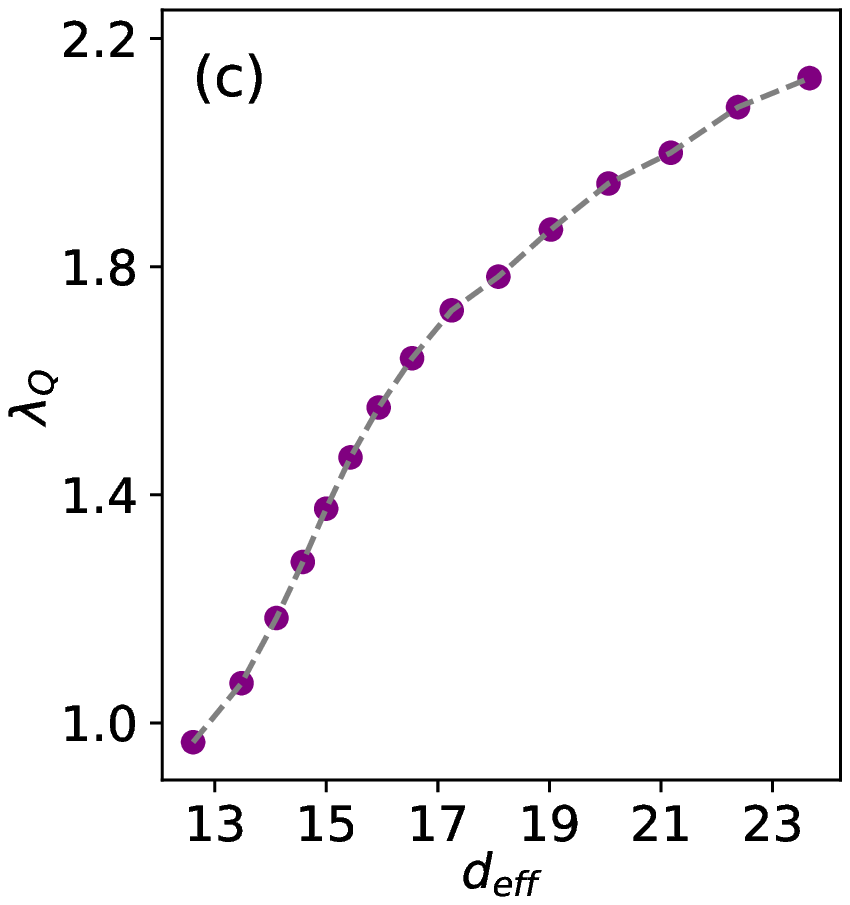}
  \includegraphics[width=0.24\textwidth]{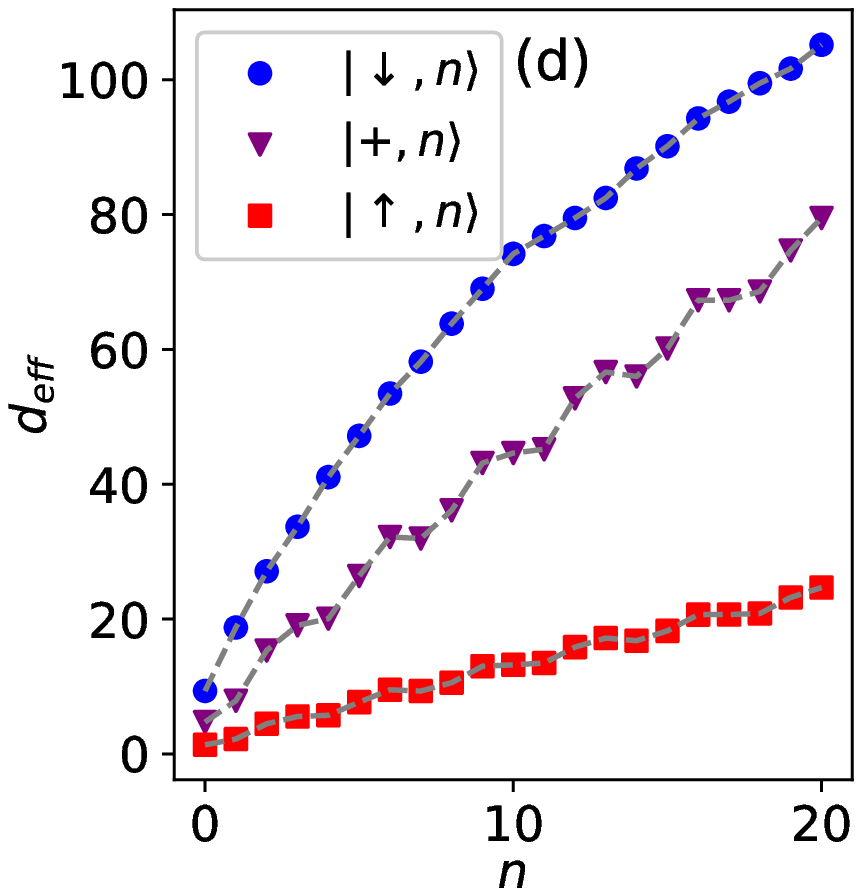}
 \includegraphics[width=0.22\textwidth]{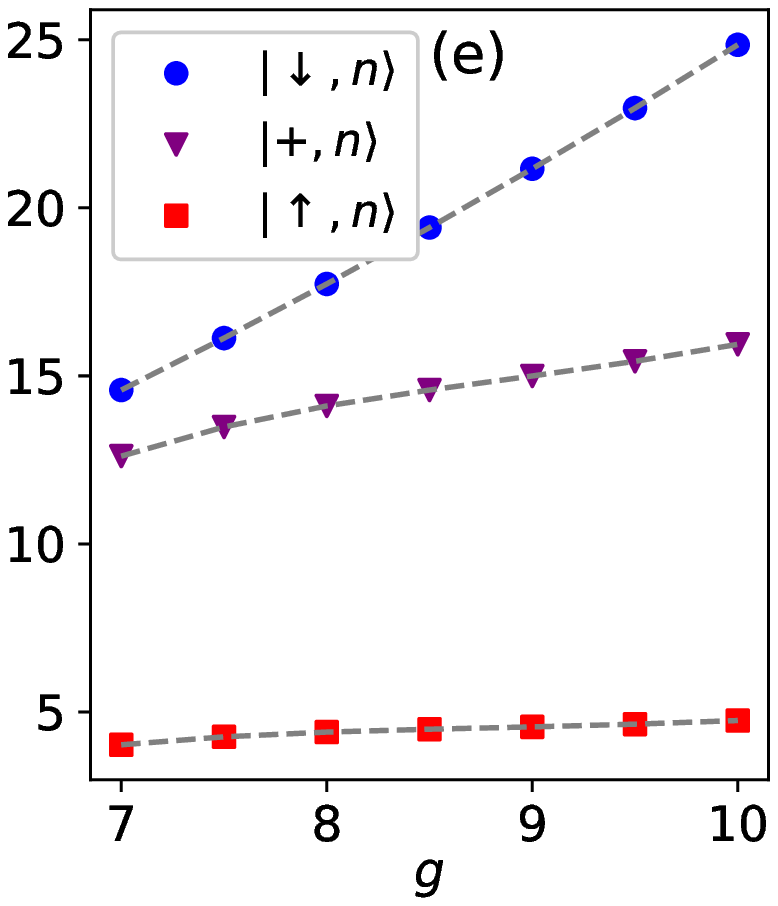}
  
\caption {(a) Time-evolution of the spin populations from $\left|\psi_0 \right\rangle = \left|\downarrow,20\right\rangle$ compared to the DE (solid lines) and ME (dashed lines) predictions in the superradiant phase for $\omega\Delta = 100$, $\eta = 200$, $g = 15$. (b) Long-time average of the spin populations (red triangles and purple stars) and mean boson number (blue dots) for various initial states and $\eta = 100$, $g = 10$. (c) Quantum Lyapunov exponent as a function of $d_{\rm eff}$ through varying $g$, $\omega \Delta = 100$, $\eta = 200$, $\left|\psi_0 \right\rangle = \left| + ,5\right\rangle$. (d) Effective dimension of the time-averaged density operator $d_{\rm eff}$ as a function of $n$ for $\eta = 200$, $g = 15$. (e)  The same but as function of $g$ for $n = 5$.  }
  \label{fig:thermal}
\end{figure}

In Fig. \ref{fig:fockstatesDE} we show that (\ref{longtimeavg}) holds for the QR model. We have chosen our observables to be the projectors onto the bosonic Fock states $\hat{P}(n) = | n \rangle \langle n|$ and $\sigma_z$ eigenstates $\hat{P}(s) = \left| s \right\rangle \left\langle s\right|$, ($s=\uparrow,\downarrow$). To find the long-time average value we use $\bar{\langle P \rangle} = \frac{1}{\Delta t} \int_t^{t +\Delta t} d\tau {\rm Tr} [ \hat{P} \hat{\rho}(\tau)]$ for all of the respective projector operators, where $\hat{\rho}(t)$ is the density matrix of the system.

The DE prediction depends on the initial state of the system through the amplitudes $c_k$, however, agreement between the DE and ME predictions implies a thermodynamical universality, namely the (averaged) relaxation value of an observable should only depend on the initial energy, and should hold true for a variety of initial states of the same energy \cite{Rigol2008}. We test this for the QR model for initial states of the type $| \psi_0 \rangle = (\left|\downarrow \right\rangle + e^{i\phi} \left|\uparrow\right\rangle)|0\rangle/\sqrt{2}$ which have the same $E_0$ for any value of $\phi$. The results presented in Fig. \ref{fig:thermal}(b) show that (\ref{longtimeavg}) leads to such universality, i.e. we have that $\sum_{k} |c_k|^2 O_{kk} = \langle \bar{O} \rangle_{E_{0}}$.
In Fig. \ref{fig:thermal}(a) we compare $\langle \bar{O} \rangle_{E_{0}}$ with ${\langle \hat{O} \rangle}_{\rm ME} (E_0) $, as well as the non-averaged time-evolution $\langle \psi(t) | \hat{O} | \psi(t) \rangle$ for $\hat{O} = \{\hat{P}(\uparrow), \hat{P}(\downarrow)\}$. The microcanonical energy shell is chosen with robustness in mind, meaning that the ME prediction gives nearly the same result regardless of small fluctuations around the value of $\delta E$.  This is in accordance with the implication of the ETH that the expectation values $O_{kk}$ of $\hat{O}$ for states $|E_k\rangle$ inside the energy shell are nearly independent of $k$. 
However, we see that in the general case the DE and ME averages do not agree, despite the aforementioned universality. 

Furthermore, we investigate the condition of equilibration of the spin system which requires the effective dimension of the time-averaged density matrix defined by $d_{\rm eff}=(\sum_{k}|c_{k}|^{4})^{-1}$ to be much larger than $d^{2}_{\rm s}$ ($d_{\rm eff}\gg d^{2}_{\rm s}$) where $d_{\rm s}=2$ is the spin system dimension \cite{Linden2009,Gogolin2011}. This condition ensures that the initial state is composed of a large number of energy eigenstates so that the bosonic degree of freedom acts as an effective bath coupled to the spin. In Figs. \ref{fig:thermal}(c) and \ref{fig:thermal}(d) we show $d_{\rm eff}$ for different initial spin and Fock states. We see that for all initial states $d_{\rm eff}$ increases with the number of bosons $n$ \cite{Clos16}.
This leads to suppression of the temporal fluctuations and hence equilibration of the spin obsevable which remains close to its time average as is shown in Fig. \ref{fig:thermal}(a).
Finally, we note that for large effective dimension $d_{\rm eff}$ the initial local information is spread between large number of eigenstates which increases $\lambda_{Q}(g,\omega,\Delta)$ and thus makes the QR system more chaotic, see Fig. \ref{fig:thermal}(c).


\section{Summary}\label{SM} 
We have shown that the critical QR model exhibits signatures of quantum chaos in the superradiant phase that become more apparent as we approach the effective thermodynamic limit $\eta \rightarrow \infty$. This is most clearly seen in the behaviour of the FOTOC which also quantifies chaos via the quantum Lyuapunov exponent. Furthermore, we investigate the equilibration of the spin degree of freedom in the QR model. We see that the system doesn't thermalize in general, but exhibits relaxation due to dephasing, characterized by the long-time average of observables that can be described using the diagonal density ensemble. We also have shown that the effective dimension of the time-averaged density matrix can be much larger than the spin system dimension which leads to equilibration of the spin observables.

\section*{Acknowledgments} 

We thank W. Li for useful discussion. A. V. K. and P. A. I. acknowledges support by the ERyQSenS project, Bulgarian Science Fund Grant No. DO02/3. D.P. acknowledges support from Spanish  project  PGC2018-094792-B-100(MCIU/AEI/FEDER, EU).


\appendix

\begin{section}{Level-spacing distribution of the quantum Rabi model}\label{QRLSD}

The quantum Rabi Hamiltonian
\begin{equation}
\hat{H}_{\rm QR}= \omega \hat{a}^\dag \hat{a} + \frac{\Delta}{2} \sigma_z + g \sigma_x (\hat{a}^\dag + \hat{a}), \label{AppRabiHamiltonian}
\end{equation}
possesses a parity symmetry generated by the operator 
\begin{equation}
\hat{\Pi}=\hat{a}^{\dag}\hat{a}+\frac{1}{2}(\sigma_{z}+1),
\end{equation}
such that $[e^{i\pi\hat{\Pi}},\hat{H}_{\rm QR}]=0$ which divides the total Hilbert state space of (\ref{AppRabiHamiltonian}) into two subspaces characterized by a parity quantum number of $\pm 1$. We use the basis $\left|s,n\right\rangle$, where $n$ is the boson Fock space number and $s$ denotes the spin polarization along the $z-$axis ($s=\uparrow,\downarrow$), and consider the negative parity subspace $ \langle s,n\left| e^{i\pi\hat{\Pi}} \right|s,n\rangle = -1$, which is composed of states of the type $\left|\uparrow ,2k\right\rangle$ and $\left|\downarrow,2k + 1\right\rangle$ for integer $k$. An effective Hamiltonian matrix of the transition amplitudes between states of this subspace is obtained by truncating the bosonic Fock space. An illustrative example for $n^{(max)}  = 5$ is given below
\begin{widetext}
\begin{equation}
H^{-}_{\rm QR}=\begin{bmatrix}
0\omega + \frac{\Delta}{2} & \sqrt{1} g & 0 & 0 &0&0  \\
\sqrt{1} g & 1\omega - \frac{\Delta}{2} & \sqrt{2} g & 0 & 0 &0\\
0 & \sqrt{2} g & 2\omega + \frac{\Delta}{2} & \sqrt{3} g & 0 &0\\
0&0& \sqrt{3} g & 3\omega - \frac{\Delta}{2} & \sqrt{4} g & 0\\
0&0&0  & \sqrt{4} g & 4\omega + \frac{\Delta}{2}  & \sqrt{5} g\\
0&0&0 & 0 &  \sqrt{5} g &  5\omega - \frac{\Delta}{2} \\
\end{bmatrix} .  \label{matrixQR}
\end{equation}
\end{widetext}
This resulting tridiagonal matrix is then diagonalized numerically to obtain the proper level-spacing distribution for this symmetry-invariant subspace.  Note that the presence of level crossing/repulsion can be obscured by improper truncation of the bosonic Hilbert spaces resulting in insufficient eigenvalues to be considered.
\end{section}

\begin{section}{Quantum Jahn-Teller model}\label{QJTM}
We consider the quantum Jahn-Teller Hamiltonian
\begin{equation}
\hat{H}_{\rm JT}= \omega \hat{a}_{r}^\dag \hat{a}_{r} + \omega \hat{a}_{l}^\dag \hat{a}_{l} + \frac{\Delta}{2} \sigma_z + g \sigma_+ (\hat{a}_{r}^\dag + \hat{a}_{l}) + g \sigma_- (\hat{a}_{r} + \hat{a}_{l}^\dag), \label{AppJTHamiltonian}
\end{equation}
which describes a two-level system interacting with two bosonic modes, denoted here by $r$, $l$. It can be seen that it is a generalization of the quantum Rabi model by applying the transformation $\hat{a}_{r}^\dag =(\hat{a}^\dag -i \hat{b}^\dag)/\sqrt{2}$ and $\hat{a}_{l}^\dag=(\hat{a}^\dag + i \hat{b}^\dag)/\sqrt{2}$ which yields
\begin{equation}
\hat{H}_{\rm JT}= \omega \hat{a}^\dag \hat{a} + \omega \hat{b}^\dag \hat{b} + \frac{\Delta}{2} \sigma_z +   \frac{g}{\sqrt{2}} \sigma_x (\hat{a}^\dag + \hat{a}) + \frac{g}{\sqrt{2}} \sigma_y (\hat{b}^{\dag} + \hat{b}). \label{AppJTHamiltonian1}
\end{equation}
Redefining the coupling above as $ g_{\rm JT}/\sqrt{2} = g_{\rm QR} = g$ the QJT Hamiltonian is equivalent to 
\begin{equation}
\hat{H}_{\rm JT} = \hat{H}_{\rm QR} + \hat{H}_{b},
\end{equation}
where $ \hat{H}_{\rm QR}$ is the quantum Rabi model describing the interaction between the two-level system and the $a$ mode, and $\hat{H}_{b} = \omega \hat{b}^\dag \hat{b} + g \sigma_y  (\hat{b}^{\dag} + \hat{b}) $ is an additional term that introduces a new degree of freedom thus makes the resulting model non-integrable.
\begin{figure}[bp]
\includegraphics[width=0.48\textwidth]{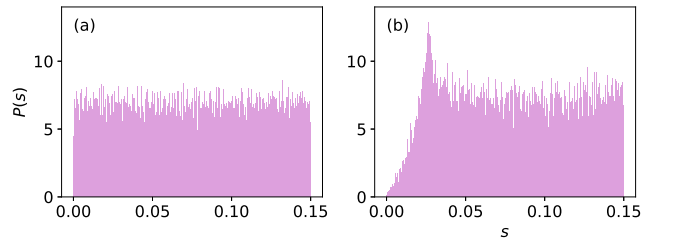}
\caption{Level-spacing distribution of the QJT model in the normal phase (left) $g = 6.07$ and the superradiant phase (right) $g=8.07$ and $\eta = 100$, $g_{\rm JT,c} = 7.07$.
\label{fig:jtlevelspacing}}
\end{figure}

\subsection{Finite size quantum phase transition in quantum Jahn-Teller model}
\subsubsection{Normal Phase}
Consider the limit $\Delta\gg \omega$ in which the spin excitations are highly suppressed. In order to find an effective description we perform a canonical transformation, namely $\hat{H}_{\rm eff}=e^{-\hat{S}}\hat{H}_{\rm JT}e^{\hat{S}}$ with $\hat{S}^{\dag}=-\hat{S}$. Using the Baker–Campbell–Hausdorff expression we get
\begin{eqnarray}
\hat{H}_{\rm eff}&=&\hat{H}_{0}+\hat{H}_{I}+\frac{1}{1!}\{[\hat{H}_{0},\hat{S}]+[\hat{H}_{\rm I},\hat{S}]\}\notag\\
&&+\frac{1}{2!}\{[[\hat{H}_{0},\hat{S}],\hat{S}]+[[\hat{H}_{\rm I},\hat{S}],\hat{S}]\}+\ldots
\end{eqnarray}
where $\hat{H}_{0}=\omega(\hat{n}_{r}+\hat{n}_{l})+(\Delta/2)\sigma_{z}$ and $\hat{H}_{I}=g\sigma_{+}(\hat{a}^{\dag}_{r}+\hat{a}_{l})+g\sigma_{-}(\hat{a}_{r}+\hat{a}^{\dag}_{l})$. Our goal is to choose $\hat{S}$ in a such a way that the terms linear in the coupling $g$ are cancelled in $\hat{H}_{\rm eff}$. This can be achieved with
\begin{equation}
    \hat{S}=\frac{g}{\omega\eta}\{\sigma_{-}(\hat{a}_{r}+\hat{a}^{\dag}_{l})-
    \sigma_{+}(\hat{a}^{\dag}_{r}+\hat{a}_{l})\}
\end{equation}
and the effective Hamiltonian becomes
\begin{eqnarray}
\hat{H}_{\rm eff}&=&\hat{H}_{0}+\frac{1}{2}[\hat{H}_{I},\hat{S}]
=\omega(\hat{n}_{r}+\hat{n}_{l})+\frac{\Delta}{2}\sigma_{z}\notag\\
&&+\frac{g^{2}}{\omega\eta}\sigma_{z}(\hat{a}^{\dag}_{r}+\hat{a}_{l})(\hat{a}_{r}+\hat{a}^{\dag}_{l}),
\end{eqnarray}
which is diagonal in the basis of $\sigma_{z}$. The Hamiltonian has a block diagonal structure with $\hat{H}_{\downarrow}$ and $\hat{H}_{\uparrow}$. Introducing position $\hat{x}_{r,l}=(\hat{a}^{\dag}_{r,l}+\hat{a}_{r,l})/\sqrt{2}$ and momentum $\hat{p}_{r,l}=i(\hat{a}^{\dag}_{r,l}-\hat{a}_{r,l})/\sqrt{2}$ operators for each bosonic mode we get
\begin{eqnarray}
\hat{H}_{\downarrow}&=&\omega\{\frac{1}{2}\left(1-\frac{g^{2}}{\Delta\omega}\right)
(\hat{p}^{2}_{r}+\hat{p}^{2}_{l}+\hat{x}^{2}_{r}+\hat{x}^{2}_{l})+\frac{g^{2}}{\Delta\omega}\hat{p}_{r}\hat{p}_{l}\notag\\
&&-\frac{g^{2}}{\Delta\omega}\hat{x}_{r}\hat{x}_{l}\},
\end{eqnarray}
where we have omitted the constant terms. Next, we perform a $\pi/4$ rotation as follows $\hat{x}_{r}=(\hat{x}_{1}-\hat{x}_{2})/\sqrt{2}$, $\hat{x}_{l}=(\hat{x}_{1}+\hat{x}_{2})/\sqrt{2}$, $\hat{p}_{r}=(\hat{p}_{1}-\hat{p}_{2})/\sqrt{2}$, and $\hat{p}_{l}=(\hat{p}_{1}+\hat{p}_{2})/\sqrt{2}$. A subsequent introduction of creation and annihilation operators for the rotated oscillator modes $\hat{d}_{1}=\sqrt{\frac{\epsilon}{2}}\hat{x}_{1}+\frac{i}{\sqrt{2\epsilon}}\hat{p}_{1}$, $\hat{d}_{2}=\frac{1}{\sqrt{2\epsilon}}\hat{x}_{2}+i\sqrt{\frac{\epsilon}{2}}\hat{p}_{2}$ yields
\begin{equation}
    \hat{H}_{\downarrow}=\omega\epsilon(\hat{d}^{\dag}_{1}\hat{d}_{1}
    +\hat{d}^{\dag}_{2}\hat{d}_{2}),
\end{equation}
where $\epsilon=\sqrt{1-(g/g_{\rm JT,c})^{2}}$ with $g_{\rm JT,c}=\sqrt{\Delta\omega/2}$ being the critical coupling. Thus in the effective thermodynamic limit $\eta\rightarrow\infty$ the normal phase with $g<g_{\rm JT,c}$ is characterized with null ground state mean bosonic excitation $\langle \hat{a}^{\dag}_{r,l}\hat{a}_{r,l}\rangle_{\rm G}/\eta=0$ and spin pointing along the $z$-axis, $\langle\sigma_{z}\rangle_{\rm G}=-1$.

\begin{figure}[tp]
\centering
\includegraphics[width=0.48\textwidth]{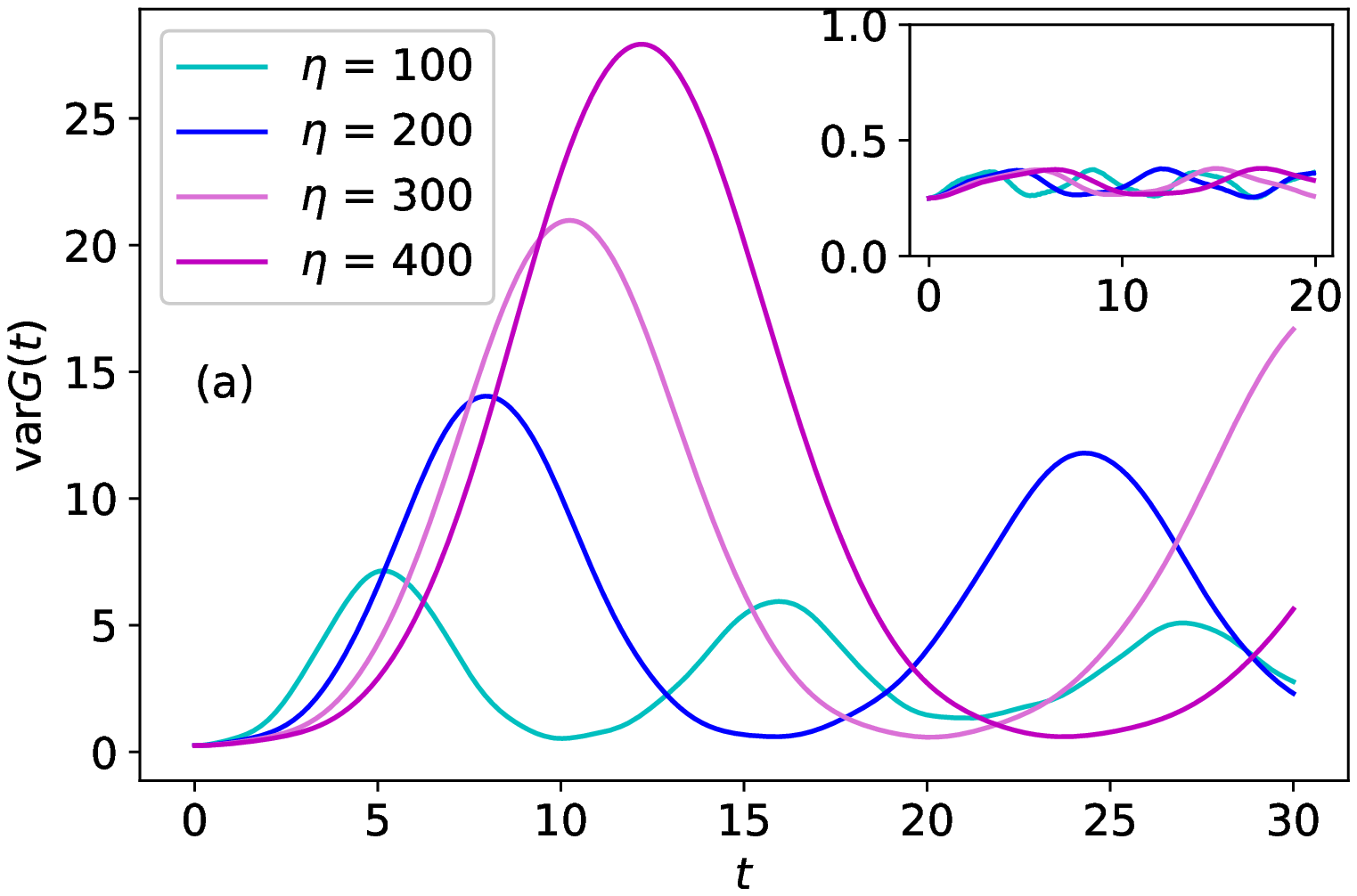}
\includegraphics[width=0.48\textwidth]{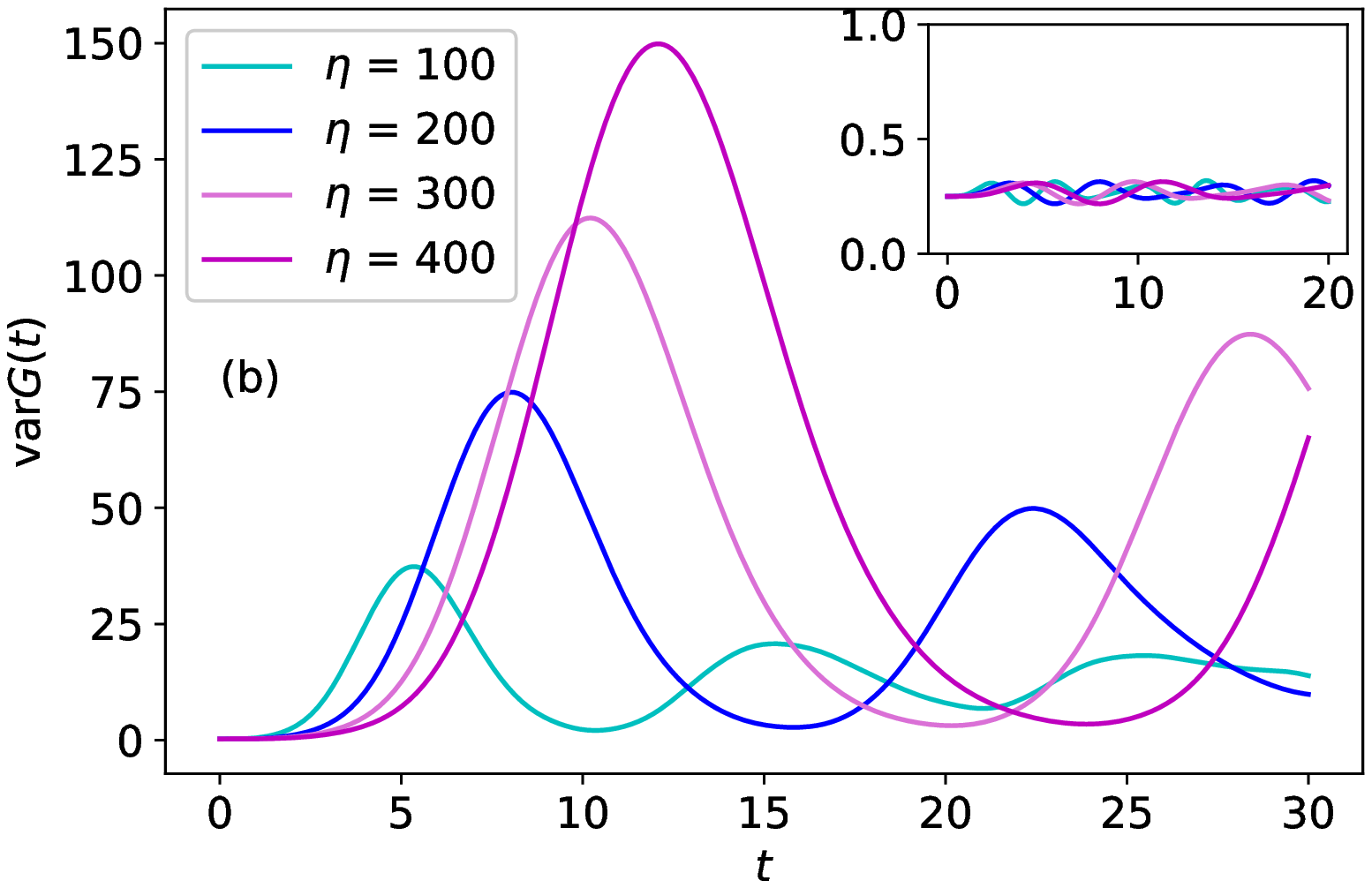}
\caption{(a) FOTOC for the QJT model in the superradiant phase (main plot) $g=9$ and normal phase (inset) $g=6$; $g_{\rm JT,c} = 7.07$. (b) FOTOC for the perturbed QR model in the superradiant phase (main plot) $g=7$ and normal phase (inset) $g=3$; $\lambda = 0.1$, $g_{\rm c} = 5$.
\label{fig:jtfotoc}}
\end{figure}

\subsubsection{Superradiant Phase}
To describe the superradiant phase, we utilize displacement operators for each bosonic mode such that the Hamiltonian $\hat{\tilde{H}}=\hat{D}^{\dag}(\alpha_{r})\hat{D}^{\dag}(\alpha_{l})\hat{H}_{\rm JT}\hat{D}(\alpha_{r})\hat{D}(\alpha_{l})$ takes the form
\begin{eqnarray}
\hat{\tilde{H}}&=&\omega(\hat{n}_{r}+\hat{n}_{l})+
\omega(\alpha_{r}\hat{a}^{\dag}_{r}+\alpha^{*}_{r}\hat{a}_{r})+\omega(\alpha_{l}\hat{a}^{\dag}_{l}+\alpha^{*}_{l}\hat{a}_{l})\notag\\
&&+\frac{\Delta}{2}\sigma_{z}+g\sigma_{+}(\hat{a}^{\dag}_{r}+\hat{a}_{l})
+g\sigma_{-}(\hat{a}_{r}+\hat{a}^{\dag}_{l})\notag\\
&&+g\sigma_{-}(\alpha^{*}_{r}+\alpha_{l})+g\sigma_{-}(\alpha_{r}+\alpha^{*}_{l}),
\end{eqnarray}
out of which we can combine the following terms that characterize solely the direction of the spin,
\begin{equation}
\hat{H}_{\rm spin}=\frac{\Delta}{2}\sigma_{z}+g\sigma_{+}
(\alpha^{*}_{r}+\alpha_{l})+g\sigma_{-}(\alpha_{r}+\alpha_{l}^{*})=\frac{\Omega}{2}\tilde{\sigma}_{z},
\end{equation}
where $\Omega=\sqrt{\Delta^{2}+4g^{2}(|\alpha_{r}|^{2}+|\alpha_{l}|^{2}+\alpha_{r}\alpha_{l}+\alpha^{*}_{r}\alpha^{*}_{l})}$, with the corresponding eigenstates $|\tilde{\downarrow}\rangle=-\cos(\theta)e^{i\phi}\left|\uparrow\right\rangle+\sin(\theta)\left|\downarrow\right\rangle$, $|\tilde{\uparrow}\rangle=\sin(\theta)\left|\uparrow\right\rangle+\cos(\theta)e^{-i\phi}\left|\downarrow\right\rangle$, where $\phi=\arg(\alpha^{*}_{r}+\alpha_{l})$ with $\cos(2\theta)=-\Delta/\Omega$. This shows that the spin basis is rotated in the superradiant phase with respect to the one in the normal phase. Transforming the raising and lowering spin operators into the new basis yields

\begin{equation}
\sigma_{+}=\frac{e^{-i\phi}}{2}\sin(2\theta)\tilde{\sigma}_{z}+\sin^{2}(\theta)\tilde{\sigma}_{+}-e^{-2i\phi}\cos^{2}(\theta)\tilde{\sigma}_{-}.    
\end{equation}
The Hamiltonian becomes
\begin{eqnarray}
\hat{\tilde{H}}&=&\omega(\hat{n}_{r}+\hat{n}_{l})+
\omega(\alpha_{r}\hat{a}^{\dag}_{r}+\alpha^{*}_{r}\hat{a}_{r}+\alpha_{l}\hat{a}^{\dag}_{l}+\alpha^{*}_{l}\hat{a}_{l})+\frac{\Omega}{2}\tilde{\sigma}_{z}\notag\\
&&+g\{\left(\frac{e^{-i\phi}}{2}\sin(2\theta)\tilde{\sigma}_{z}+\sin^{2}(\theta)\tilde{\sigma}_{+}-e^{-2i\phi}\cos^{2}(\theta)\tilde{\sigma}_{-}\right)\notag\\
&&\times(\hat{a}^{\dag}_{r}+\hat{a}_{l})
+{\rm H.c.}\}\label{Hss}
\end{eqnarray}
The displacement parameters $\alpha_{r,l}$ can be found by the condition that all terms linear in the bosonic operators in (\ref{Hss}) are cancelled. Projecting these terms onto the spin state $|\tilde{\downarrow}\rangle$ we obtain
\begin{equation}
\omega\alpha_{r}-g\frac{e^{-i\phi}}{2}\sin(2\theta)=0,\quad   \omega\alpha_{l}-g\frac{e^{i\phi}}{2}\sin(2\theta)=0. 
\end{equation}
We find that for $g<g_{\rm JT,c}$ the parameters are $\alpha_{r}=\alpha_{l}=0$ and respectively for $g>g_{\rm JT,c}$ we have
\begin{equation}
    |\alpha^{*}_{r}+\alpha_{l}|=\sqrt{\frac{\eta}{2\lambda^{2}}(\lambda^{4}-1)},
\end{equation}
with $\lambda=g/g_{\rm JT,c}$. Using this the Hamiltonian becomes
\begin{eqnarray}
\hat{\tilde{H}}&=&\omega(\hat{n}_{r}+\hat{n}_{l})+\frac{\Omega}{2}\tilde{\sigma}_{z}+
g\{\tilde{\sigma}_{+}(\sin^{2}(\theta)(\hat{a}^{\dag}_{r}+\hat{a}_{l})\notag\\
&&-e^{2i\phi}\cos^{2}(\theta)(\hat{a}_{r}+\hat{a}^{\dag}_{l}))+{\rm H.c.}\}
\end{eqnarray}
In the limit $\eta\rightarrow\infty$ we can perform a canonical transformation with

\begin{equation}
    \hat{S}=\frac{g}{\Omega}\{\tilde{\sigma}_{-}(\sin^{2}(\theta)(\hat{a}_{r}+\hat{a}^{\dag}_{l})-e^{-2i\phi}\cos^{2}(\theta)(\hat{a}^{\dag}_{r}+\hat{a}_{l}))-{\rm H.c.}\},
\end{equation}
such that the effective Hamiltonian, which is projected on the state $|\tilde{\downarrow}\rangle$ becomes
\begin{eqnarray}
\hat{\tilde{H}}_{\tilde{\downarrow}}&=&\omega(\hat{n}_{r}+\hat{n}_{l})-\frac{g^{2}}{\Omega}
\{(\cos^{4}(\theta)+\sin^{4}(\theta))(\hat{a}^{\dag}_{r}+\hat{a}_{l})\notag\\
&&\times(\hat{a}_{r}+\hat{a}^{\dag}_{l})-\sin^{2}(\theta)\cos^{2}(\theta)(e^{2i\phi}(\hat{a}_{r}+\hat{a}^{\dag}_{l})^{2}\notag\\
&&+e^{-2i\phi}(\hat{a}^{\dag}_{r}+\hat{a}_{l})^{2})\}.
\end{eqnarray}
Introducing position and momentum operators via the relations $\hat{a}_{r}=e^{-i\phi}(\hat{x}_{1}-\hat{x}_{2}+i(\hat{p}_{1}-\hat{p}_{2}))/2$ and $\hat{a}_{l}=e^{i\phi}(\hat{x}_{1}+\hat{x}_{2}+i(\hat{p}_{1}+\hat{p}_{2}))/2$ we obtain
\begin{eqnarray}
\hat{\tilde{H}}_{\tilde{\downarrow}}&=&\omega\{\frac{\hat{p}^{2}_{1}}{2}+\frac{\hat{p}^{2}_{2}}{2}\left(1-\frac{2g^{2}}{\omega\Omega}\right)+\frac{x^{2}_{1}}{2}
\left(1-\frac{2g^{2}}{\omega\Omega}\cos^{2}(2\theta)\right)\notag\\
&&+\frac{\hat{x}^{2}_{2}}{2}\}.
\end{eqnarray}
We find one mode $\tilde{\epsilon}_{1}=0$ which corresponds to a free mode. The latter is the Goldstone mode related to the breaking of the continuous U(1) symmetry. The second mode is $\tilde{\epsilon}_{2}=\sqrt{1-\frac{1}{\lambda^{4}}}$ which is defined for $g>g_{\rm JT,c}$. The superradiant phase is characterized with spin orientation $\langle\sigma_{z}\rangle_{\rm G}=\cos(2\theta)$ and mean bosonic excitation $\lim_{\eta\rightarrow \infty}\langle (\hat{a}^{\dag}_{r}+\hat{a}_{l})(\hat{a}_{r}+\hat{a}^{\dag}_{l})\rangle/\eta=\frac{\lambda^{4}-1}{2\lambda^{2}}$.

\begin{figure}[bp]
\centering
\includegraphics[width=0.48\textwidth]{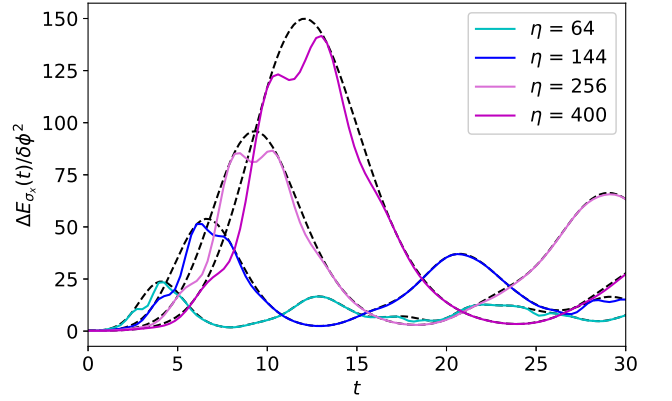}
\caption{Echo signal for $\sigma_x$ in the QR model under imperfect time reversal in the superradiant phase (solid lines). Dashed black lines correspond to the FOTOC from the main text. The parameters are set to $g = 7$, and various $\eta$.
\label{fig:echoofsigmax}}
\end{figure}

\subsection{Signatures of chaos in the quantum Jahn-Teller model}

The QJT Hamiltonian (\ref{AppJTHamiltonian}) has a continuous $U(1)$ symmetry given by the operator $\hat{C} = \hat{a}_{l}^\dag \hat{a}_{l} - \hat{a}^\dag_r \hat{a}_{r} + \frac{1}{2}\hat{\sigma}_z$ which separates the total state space spanned by the basis  $|s, n_r , n_l \rangle$, where ($s=\uparrow,\downarrow$), $\hat{a}^\dag_{r,l} \hat{a}_{r,l} |s, n_r , n_l \rangle = n_{r,l} |s, n_r , n_l \rangle$, into invariant subspaces for every half-integer eigenvalue $c$ of $\hat{C} |s, n_r , n_l \rangle = c |s, n_r , n_l \rangle$. Here we investigate the case $c = \frac{3}{2}$ which fixes the symmetry invariant subspace containing states of the type $\left|\uparrow,n,n+1\right\rangle$ and $\left|\downarrow, m, m+2\right\rangle$ for integer $m,n$. Similarly to the QR model, this procedure yields a tridiagonal matrix for the QJT Hamiltonian (\ref{AppJTHamiltonian}) which is then numerically diagonalized to find the level-spacing distribution. An example for such a matrix for a bosonic Fock space truncated at $n_l^{(max)} =4$, $n_r^{(max)} =3$ and $c=\frac{3}{2}$ is given by

\begin{widetext}
\begin{equation}
H_{\rm JT}^{(\frac{3}{2})} = \begin{bmatrix}
1\omega + \frac{\Delta}{2} & \sqrt{2} g & 0 & 0 &0&0 &0 \\
\sqrt{2} g & 2\omega - \frac{\Delta}{2} & \sqrt{1} g & 0 & 0 &0 &0\\
0 & \sqrt{1} g & 3\omega + \frac{\Delta}{2} & \sqrt{3} g & 0 & 0&0\\
0&0& \sqrt{3} g & 4\omega - \frac{\Delta}{2} & \sqrt{2} g & 0 &0\\
0&0&0  & \sqrt{2} g & 5\omega + \frac{\Delta}{2}  & \sqrt{4} g &0\\
0&0&0 & 0 &  \sqrt{4} g &  6\omega - \frac{\Delta}{2}  &  \sqrt{3} g\\
0&0&0 & 0 & 0&\sqrt{3} g &7\omega + \frac{\Delta}{2} \\
\end{bmatrix}  . \label{matrixJT}
\end{equation}
\end{widetext}

As in the QR model, we observe neither Poissonian nor Wigner-Dyson distribution in both phases of the system, however, focusing on a smaller scale for the nearest-neighbour energy difference, Fig. \ref{fig:jtlevelspacing} shows that level-crossings are present in the normal phase, while level-repulsions characterize the supperradiant phase.

We further investigate signatures of chaos in the QJT model using the FOTOC as defined in the main text $F(t) = \langle \hat{W}_G^\dag (t) \hat{V}^\dag \hat{W}_G(t) \hat{V} \rangle$, where  $\hat{W}_G(t) = e^{i\hat{H}_{\rm JT}t} \hat{W}_G e^{-i\hat{H}_{\rm JT}t}$, $\hat{W}_G = e^{i\delta\phi \hat{G}}$. We plot the variance of the operator $\hat{G}=(\hat{a}_{r}^\dag + \hat{a}_{r})/2$ in Fig. \ref{fig:jtfotoc}(a) and find that it exhibits similar behaviour to that of the FOTOC for the QR model, namely we observe an initial exponential growth in the superradiant phase that is related to a quantum Lyapunov exponent, followed by saturation and long-time oscillations. In the normal phase the FOTOC oscillates with amplitude independent of the thermodynamical parameter $\eta$ for all $t$.

\end{section}
\begin{section}{Echo signal of $\sigma_x$}\label{echo}

We can further showcase the sensitivity to small perturbations of our chaotic system by choosing a different $\hat{V}$ operator for the FOTOC, and making use of the explicit connection between the FOTOC and the Loschmidt echo signal $E_{V} (t) = \langle\psi_{0}| \hat{U}_{\delta\phi}^\dag \hat{V} \hat{U}_{\delta\phi}|\psi(0)\rangle$, where $\hat{U}_{\delta\phi} = e^{i\hat{H}t}e^{i\delta\phi \hat{G}}e^{-i\hat{H}t}$. The divergence from the perfect echo is given by 
\begin{eqnarray}
\Delta E_{V}&=&\langle\psi_{0}|\hat{V}|\psi_{0}\rangle-E_{V}(t)\notag\\
&&=  \frac{\delta\phi^2}{2} \langle \psi_{0}| [\hat{G}(t), [\hat{G}(t), \hat{V}]] |\psi_{0}\rangle+ O(\delta\phi^3).
\end{eqnarray}

\begin{figure}[tbp]
\centering
\includegraphics[width=0.48\textwidth]{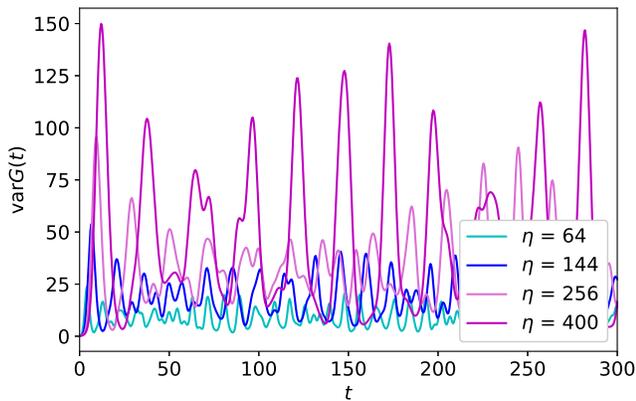}
\caption{Long-time behaviour of the FOTOC as defined in the main text for  $g = 7$, $g_{\rm c}=5$ and various $\eta$.
\label{fig:longtime}}
\end{figure}

This quantity corresponds to the variance of $\hat{G}$ in the case of $\hat{V} = |\psi_{0}\rangle  \langle \psi_{0}|$. 
Setting $ \hat{V} = \sigma_x$, we plot $\Delta E_{\sigma_x} / \delta\phi^2$ in Fig. \ref{fig:echoofsigmax}. We once again observe initial exponential growth in the echo signal with a comparable quantum Lyapunov exponent to the one extracted from the FOTOC. Furthermore, we see that the echo signal, whilst following a similar pattern of growth to the FOTOC, displays some oscillations throughout its evolution, and seems to always be bounded by the corresponding value of the FOTOC.
\end{section}

\begin{section}{Long-time behaviour of the FOTOC}\label{long}

In the main text we put emphasis on the initial exponential growth of the FOTOC prior to the scrambling time $t^*$ from which we extract the quantum Lyapunov exponent. Fig. \ref{fig:longtime} shows the long-time behaviour beyond $t^*$, namely we observe oscillations, followed by periodic maximal saturation of the FOTOC. This behaviour seems to persist regardless of the time period, hence the FOTOC does not reach true saturation. Such a repeated near-saturation peaks are likely due to the finite-size of the model, as collective systems such as the Dicke model that may also exhibit similar oscillatory behaviour do not reach an amplitude comparable to that of the initial peak.

\end{section}

\end{document}